\pdfoutput=1

\documentclass[11pt]{article}

\usepackage[final]{acl}

\usepackage{times}
\usepackage{latexsym}

\usepackage[T1]{fontenc}

\usepackage[utf8]{inputenc}

\usepackage{microtype}

\usepackage{inconsolata}

\usepackage{graphicx}

\usepackage{amssymb}
\usepackage{amsmath}

\usepackage{booktabs}
\usepackage{multirow}
\usepackage{multirow}
\usepackage{multicol}
\usepackage{booktabs}
\usepackage{arydshln}
\usepackage{color}
\usepackage{makecell}
\usepackage{subcaption}
\usepackage{tabularx}
\usepackage{caption}
\usepackage{pifont}
\usepackage{colortbl}
\definecolor{mygray}{gray}{.9}
\usepackage{makecell}
\renewcommand{\arraystretch}{0.93}
\usepackage[most]{tcolorbox}
\usepackage{float}
\usepackage{xspace}

\usepackage{enumitem}

\tcbset{
  promptbox/.style={
    width=0.95\textwidth,
    colback=white,
    colframe=teal!80!black,
    colbacktitle=teal!50,
    fonttitle=\bfseries\color{white},
    coltitle=white,
    boxrule=0.5mm,
    arc=4mm,
    top=10pt,
    bottom=10pt,
    left=10pt,
    right=10pt,
    enhanced,
    center,
    attach boxed title to top left={xshift=5mm, yshift=-3mm},
    boxed title style={arc=3mm, colframe=teal!80!black},
    title filled,
    drop shadow={xshift=1mm, yshift=-1mm, fill=teal!50!black}
  }
}

\newtcolorbox{PromptBox}[2][]{promptbox, title=#2, #1}

\newcommand{\aka}{a.k.a~}

\newcommand{\secref}[1]{Section~\ref{#1}}

\newcommand{\figref}[1]{Figure~\ref{#1}}
\newcommand{\tableref}[1]{Table~\ref{#1}}

\newcommand{\ours}{NGM-RAG}

\title{\ours: Neural Graph Matching based Retrieval-Augmented Generation}

\author{
 \textbf{Guo Chen\textsuperscript{1,\dag}},
 \textbf{Ziwen Li\textsuperscript{1,\dag}},
 \textbf{Maolin Zheng\textsuperscript{1}},
 \textbf{Hao Gao\textsuperscript{2}},
\\
 \textbf{Junjie Huang\textsuperscript{1,*}},
 \textbf{Tao Jia\textsuperscript{1}}
\\
\\
 \textsuperscript{1}College of Computer and Information Science, Southwest University, China
\\
 \textsuperscript{2}Beijing Institute of Control Engineering, China
\\
 \small{
   \texttt{\{cg1281838223, pique0202, zml778922461\}@email.swu.edu.cn}
 }
\\
 \small{
   \texttt{goleey@163.com, junjiehuang@swu.edu.cn, tjia@swu.edu.cn}
 }
\\
 \small{
   \textsuperscript{\dag}Equal contribution.
   \quad
   \textsuperscript{*}Corresponding author.
 }
}

\begin{document}

\maketitle

\begin{abstract}

Retrieval-Augmented Generation (RAG) significantly enhances the ability of Large Language Models (LLMs) to provide accurate and contextually relevant answers by dynamically integrating external databases. 
However, traditional RAG methods are primarily constrained by their reliance on text-based retrieval strategies, which often struggle with complex questions requiring multi-hop reasoning. 
To address this limitation, we introduce \underline{N}eural \underline{G}raph \underline{M}atching based \underline{R}etrieval-\underline{A}ugmented \underline{G}eneration (NGM-RAG), a novel framework that leverages graph structures to effectively capture and utilize relational knowledge for improved retrieval and answer generation. 
NGM-RAG explicitly incorporates graph construction, graph matching, and answer generation into a unified process. Within this framework, we propose a neural graph matching approach that combines text-based matching with Graph Neural Networks (GNNs). 
By employing an adaptive weighting strategy, NGM-RAG efficiently integrates multiple matching methods to select the most relevant contextual node information for answer generation.
Experimental results on multi-hop question answering and long-context summarization tasks demonstrate that our NGM-RAG model achieves superior performance compared to both traditional NaiveRAG methods and state-of-the-art graph-enhanced approaches such as GraphRAG and LightRAG.
\end{abstract}

\section{Introduction}

Retrieval-Augmented Generation (RAG) is an innovative approach that enables Large Language Models (LLMs) to generate more accurate and contextual responses, significantly improving their practicality in real-world applications~\cite{es2024ragas,salemi2024evaluating}. 
Compared to traditional large model Supervised Fine-Tuning (SFT) and In Context Learning (ICL)~\citep{taori2023alpaca,zhang2023instruction} methods, RAG can adapt to specific domain knowledge to ensure that the information provided is not only relevant but also tailored to the user's needs. 
In addition, RAG can dynamically update private data, so it has received widespread attention from academia and industry~\cite{gao2023retrieval}.

Although RAG effectively leverages external and up-to-date databases, its performance heavily depends on the accuracy of the retrieved context~\cite{zhao2024retrieval}. 
However, in real-world scenarios, many questions require multi-hop reasoning to answer correctly~\cite{yang2024large}. 
Traditional text retrieval methods, such as sparse retrieval~\cite{robertson2009probabilistic} or dense retrieval~\cite{karpukhin2020dense}, cannot effectively support LLMs in answering complex questions. 

\begin{figure}
    \centering
    \includegraphics[width=\linewidth]{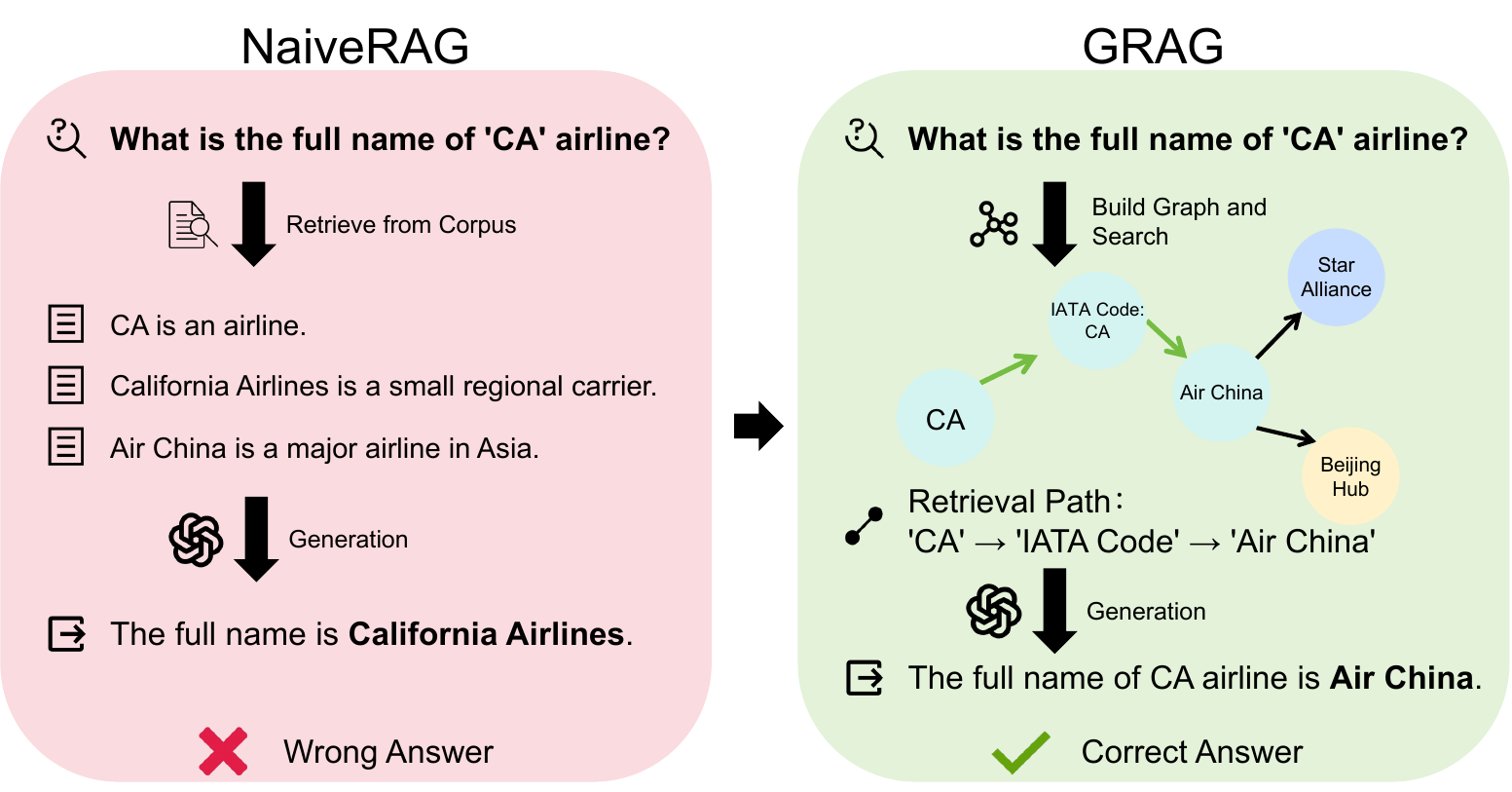}
    \caption{An example question that NaiveRAG may answer incorrectly due to inaccurate retrieval, whereas GRAG can provide the correct answer through knowledge graph relationships.}
    \label{fig:fig-1}
    \vspace{-10pt}
\end{figure}

For example, as shown in \figref{fig:fig-1}, given a question like \texttt{What is the full name of the `CA' airline?}, traditional RAG retrieves contexts related to `CA', but these contexts lack semantic connections. 
As a result, the LLM generates an incorrect answer.
To better organize and manage knowledge, graph structures (with their intrinsic nature of ``nodes connected by edges", encoding rich, heterogeneous, and relational information) are integrated into the RAG framework~\cite{han2024retrieval}.
By modeling knowledge graphs and performing graph search, ``CA" can be associated with ``Air China" via its ``IATA code", enabling the retrieval of the correct answer.

Although graph structures have already been explored in RAG, such as GraphRAG~\cite{edge2024local} and LightRAG~\cite{guo2024lightrag}, effectively leveraging the relational information of knowledge graphs—namely, graph structures—still lacks a clear solution. 
In this paper, we first provide a formal definition of Graph Retrieval-Augmented Generation (GRAG), and then propose a novel Neural Graph Matching RAG (NGM-RAG) framework comprising graph construction, graph matching, and answer generation. 
For the graph matching, in addition to identifying relevant nodes via text similarity, we further propose a Neural Graph Matching method based on Graph Neural Networks (GNNs).
Our Neural Graph Matching method can be applied to parameter-free homogeneous graph neural networks, such as LightGCN~\cite{he2020lightgcn}. 
It can also be adapted to heterogeneous graph neural networks, such as GINE~\cite{Hu2019StrategiesFP}, by introducing supervision signals.
By applying an adaptive weighting strategy, different matching algorithms can be effectively integrated. We then incorporate the most relevant information into the context to perform retrieval-augmented generation.
Empirical results on multiple multi-hop question answering and long-context summarization tasks demonstrate that our proposed model outperforms both NaiveRAG and state-of-the-art GRAG methods, such as GraphRAG and LightRAG.

We summarize our contributions as follows:
\begin{itemize}
    \item We present a formal definition of GRAG, and based on this formalization, propose a Neural Graph Matching-based RAG framework, named NGM-RAG.
    \item NGM-RAG consists of three components: graph construction, graph matching, and answer generation. For graph matching, we integrate direct matching, text similarity, and neural graph matching algorithms into an adaptive framework that combines different matching functions.
    \item NGM-RAG achieves superior performance compared to several baselines on multiple multi-hop question answering and long-context summarization tasks.
\end{itemize}

\section{Related Work}

\subsection{Graph Retrieval Augmented Generation}
Retrieval-Augmented Generation (RAG) enhances language models by integrating external knowledge during generation, typically using semantic or lexical similarity for retrieval. 
However, traditional RAG methods are limited in effectively utilizing structured, relational data~\cite{han2024retrieval}.
Graph Retrieval Augmented Generation (GRAG) addresses this limitation by incorporating graph-structured data into the retrieval process. Graphs naturally encode rich relational information, making them ideal for enhancing retrieval~\cite{han2024retrieval}.
GraphRAG utilizes specialized techniques such as Graph Neural Networks (GNNs)~\cite{kipf2016semi} and graph traversal algorithms~\cite{wang2024knowledge}, allowing retrieval based on relational paths rather than just embeddings~\cite{mavromatis2024gnn}.
Overall, GraphRAG effectively captures structural nuances, improving the quality and accuracy of language models, especially in complex reasoning and domain-specific tasks~\cite{mavromatis2024gnn,hu2024grag}.

\subsection{Neural Graph Matching}
\label{sec:neural-graph-matching}
As an efficient neural-based method for modeling graph-structured data, Graph Neural Networks (GNNs) have been widely used in fields including knowledge graphs~\cite{zhang2024knowledge}, and recommendation systems~\cite{he2020lightgcn}.

GNNs were first employed to model homogeneous graphs, in which all nodes belong to the same entity type~\cite{kipf2016semi,Velickovic2017GraphAN,Hamilton2017InductiveRL}. 
When extended to knowledge graphs (i.e., heterogeneous graphs), researchers have adopted modeling strategies typified by the Relational Graph Convolutional Network (R-GCN)~\cite{Schlichtkrull2017ModelingRD}. 
Unlike a plain GCN (and its simplified variant SGCN~\cite{wu2019simplifying,he2020lightgcn}, which can forego graph-level parameters and rely solely on node embeddings), heterogeneous settings must capture complex relation patterns and mappings across diverse node types; consequently, such models generally incorporate additional trainable parameters~\cite{phan2022hetergraphlongsum}.
Due to their powerful capability to model graph structures, Graph Neural Networks (GNNs) have also been widely applied to graph matching problems, enabling neural graph matching~\cite{lou2020neural,roy2022interpretable}.


\clearpage
\section{Problem Definition}
\label{sec:problem-definition}
Retrieval-Augmented Generation (RAG) system ~\citep{guo2024lightrag} can be defined as the following framework, denoted as $\mathcal{F}$:
\begin{equation}
\begin{aligned}
    \mathcal{F} &= (\mathcal{M}_{\text{Gen}}, \mathcal{M}_{\text{Ret}}), \\
    \mathcal{F}(q; D) &= \mathcal{M}_{\text{Gen}}\big(q,  \mathcal{M}_{\text{Ret}} (q, D)\big),
\end{aligned}
\end{equation}
where $\mathcal{M}_{\text{Gen}}$ is the Generation Modules, $\mathcal{M}_{\text{Ret}}$ is the Retriever Modules, $q$ is the user input query, $D$ is the external Database for query (\aka Corpus).

In general, the Generation Module $\mathcal{M}_{\text{Gen}}$ is an LLM that produces high-quality answers based on the user input query and the most relevant information obtained by Retriever Modules $\mathcal{M}_{\text{Ret}}$:
\begin{equation}
    \mathcal{M}_{\text{Ret}} \big(q, D) = \mathcal{R}(q, \mathcal{I}(D)\big),
\end{equation}
where Retrieval Module $\mathcal{M}_{\text{Ret}}$ usually consists of two parts: an Indexer $\mathcal{I}$ for external database and a Retriever $\mathcal{R}$ for matching query and indexed data.

For Retriever Module $\mathcal{M}_{\text{R}}$, a common approach is to embed text into vector representations and do semantic retrieval~\cite{gao2023retrieval}. 

In our study, we focus on Graph Retrieval-Augmented Generation (GRAG), which introduces graph structure (e.g., knowledge graph) into Retrieval Module $\mathcal{M}^{G}_{\text{R}}$, including GRAG Retriever $\mathcal{R}^G$ and Indexer $\mathcal{I}^G$, as follows:
\begin{equation}
\begin{aligned}
    \mathcal{M}^{G}_{\text{Ret}} &= \mathcal{R}^{G} (G_{\text{query}}, G_{\text{target}}), \\
    G_{\text{query}} & = \mathcal{I}^G (q), G_{\text{target}} = \mathcal{I}^{G}(D), 
\end{aligned}
\end{equation}
where $G_{\text{target}}$ and $G_{\text{query}}$ are the target graph and query graph produced by Graph Indexer $\mathcal{I}^{G}$ respectivelly; $\mathcal{R}^{G}$ is a graph matching function to match the query graph and target graph.
More specifically, given a query, $\mathcal{I}^{G}$ transforms text into a knowledge graph $G=(V, E)$, where $V$ is the node set and $E$ is the edge set.
$\mathcal{R}^{G}$ performs matching between two graphs, to find the most relevent nodes $V_{\text{rel}}$ and edges $E_{\text{rel}}$ for Generation Modules $\mathcal{M}_{\text{Gen}}$, as follows:
\begin{equation}
    \mathcal{M}^G_\text{Ret}(q, D) = (V_\text{rel}, E_\text{rel}).
\end{equation}

The goal of GRAG is to design an efficient Graph Retriever Module $\mathcal{M}^{G}_{\text{Ret}}$ that has a stronger ability to solve complex and comprehensive problems than traditional RAG.
Common graph-based RAG (e.g., GraphRAG~\citep{edge2024local} and LightRAG~\citep{guo2024lightrag}) adopt the above paradigm with different designs on Graph Retriever Module.

\section{Methodology}

\begin{figure*}[t]
    \centering
    \includegraphics[width=\textwidth]{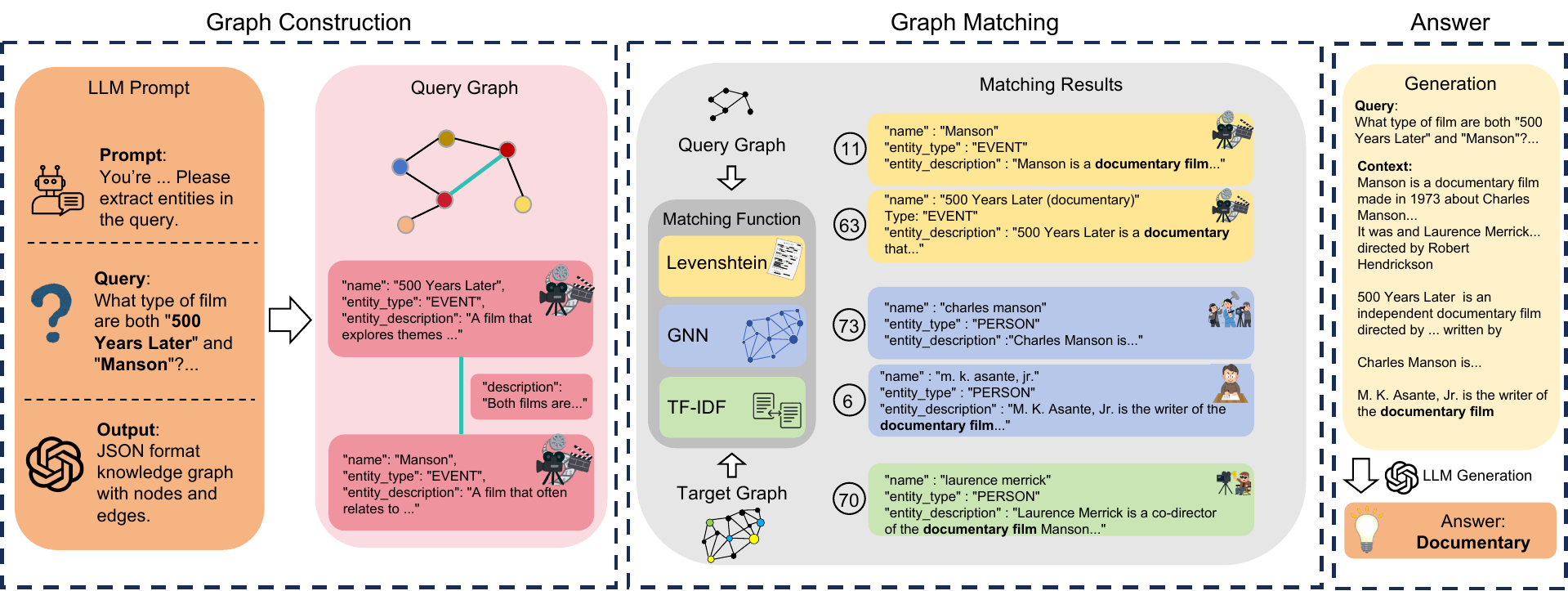}
    \caption{Overall workflow of our NGM-RAG, which consists of multiple components, including Graph Construction, Graph Matching, and Context and Answer Generation.}
    \label{fig:fig-ngm-rag}
\end{figure*}

Based on the definition of GRAG, we proposed a new \underline{R}etrieval-\underline{A}ugmented \underline{G}eneration with \underline{N}eural \underline{G}raph \underline{M}atching, named \ours~ in \figref{fig:fig-ngm-rag}.

\subsection{Graph Construction}
\label{sec:graph-construction}
Given a dataset corpus, we first leverage LLMs to build a target knowledge graph $G_{\text{target}} = (V_\text{target}, E_\text{target})$, a representation widely adopted in GRAG ~\cite{guo2024lightrag}, where $V_\text{target}$ is the node set and $E_\text{target}$ is the edge set, respectively. 
For a given the query, we construct its graph $ G_{\text{query}} $ in the same manner.
Each node and edge is annotated with its corresponding textual information, including name, type and description for a node and  description for an edge shown in \figref{fig:fig-ngm-rag}.
More detail about prompts can be found in \secref{sec:graph-contstruction}.
It is worth noting that the Graph Construction process serves for data distillation and indexing. 
Therefore,  the Graph Indexer Module in GRAG can be formed as:
\begin{equation}
    \mathcal{I}^G = \text{LLM}_\text{prompt}(q, D) \rightarrow G_{\text{target}}, G_{\text{query}},
\end{equation}
where $\text{LLM}_\text{prmopt}$ is a powerful LLM with specific prompts for knowledge graph extraction\footnote{
In this work, we used GPT-4o-mini for our method and baselines.}. 

\subsection{Graph Matching}
After completing the graph construction in \secref{sec:graph-construction}, 
We aim to identify the most relevant nodes in $G_{\text{target}}$ with respect to $G_{\text{query}}$. 
This is a graph matching problem. For each node $v \in G_{\text{query}}$, we aim to find the top $K$ most similar nodes in $G_{\text{target}}$ based on the matching score $f(v, u)$. Formally,
\begin{equation}
    \mathcal{N}_K(v) = \operatorname{TopK}_{u \in V_{\text{target}}} f(v, u),
\end{equation}
where $f(v, u)$ denotes the matching score between node $v$ and node $u$, $\mathcal{N}_K(v)$ denotes the set of $K$ nodes in $G_{\text{target}}$ that have the highest matching scores with $v$.
We design three different approaches to compute the matching score: \textit{Direct Matching}, \textit{Text Similarity}, and \textit{Graph Neural Matching}.

\subsubsection{Direct Matching}

Direct Matching selects target nodes whose names match those in the query graph. Since LLM-extracted node names may vary across graphs, we use the Levenshtein distance~\cite{yujian2007normalized} to compute name-level similarity:
\begin{equation}
    S_{\text{levenshtein}} = \text{Levenshtein}(v.\text{name},u.\text{name}),
\end{equation}
where $v.\text{name}$ and $u.\text{name}$ denote the names of query node $v$ and target node $u$, respectively.

\subsubsection{Text Similarity}

To capture semantic relevance beyond exact name matching, we measure the textual similarity between node descriptions. We first use TF-IDF~\cite{schutze2008introduction} as a basic formulation:
\begin{equation}
    S_{\text{tfidf}}=\text{TFIDF}(v.\text{description},u.\text{description}).
\end{equation}
In practice, we adopt BM25~\cite{robertson2009probabilistic}, which further considers term saturation and document length normalization:
\begin{equation}
\label{eq:bm25_abstract}
S_{\text{BM25}}
= \sum_{i=1}^{n} W_i \cdot R(q_i,d),
\end{equation}
where $q_i$ is the $i$-th query term, $W_i$ is its weight, and $R(q_i,d)$ measures its relevance to document $d$.

\subsubsection{Neural Graph Matching}

Direct Matching and Text Similarity mainly rely on textual signals and do not fully exploit graph topology. To incorporate structural dependencies, we use GNNs to encode nodes and compute graph-aware matching scores:
\begin{equation}
    S_{\text{gnn}} = \text{Cosine}(\text{GNN}(v), \text{GNN}(u)),
\end{equation}
where $\text{GNN}(v)$ and $\text{GNN}(u)$ are the graph-enhanced representations of nodes $v$ and $u$, respectively. We initialize node features with BGE embeddings~\cite{bge_embedding} and adopt LightGCN~\cite{he2020lightgcn} and GINE~\cite{Hu2019StrategiesFP} as representative parameter-free homogeneous and parameterized heterogeneous GNN backbones.

For LightGCN, node representations are updated by parameter-free neighborhood aggregation:
\begin{equation}
    x_v^{l} = \sum_{j\in \mathcal{N}_v}
    \frac{1}{\sqrt{|\mathcal{N}_v|}\sqrt{|\mathcal{N}_j|}}x_j^{l-1},
\end{equation}
where $\mathcal{N}_v$ and $\mathcal{N}_j$ denote the neighbors of nodes $v$ and $j$, respectively.

For GINE, the propagation rule is:
\begin{equation}
    x^{l}_v = h_{\mathbf{\Theta}} \left(
    x_v^{l-1} + \sum_{j \in \mathcal{N}_v} \mathrm{ReLU}
    (x_j^{l-1}+e_{jv}) \right),
\end{equation}
where $h_{\mathbf{\Theta}}$ is a neural network and $e_{jv}$ denotes edge attributes. Since GINE contains trainable parameters, we use node pairs retrieved by Direct Matching and Text Similarity as supervision signals, and optimize the model with contrastive loss~\cite{chen2020simple}. Let $s_i^+ = \mathrm{sim}(x_i, x_i^+)$ and $s_i^- = \mathrm{sim}(x_i, x_i^-)$.
The contrastive loss is defined as:
\begin{equation}
\label{eq:loss-function}
\mathcal{L}
= -\sum_i \log
\frac{\exp(s_i^+/\tau)}
{\exp(s_i^+/\tau) + \exp(s_i^-/\tau)}.
\end{equation}
where $x_i^+$ and $x_i^-$ are positive and negative samples for $x_i$, respectively, and $\text{sim}(\cdot)$ denotes cosine similarity.

\subsubsection{Overall Ranking}

We combine the three matching signals through adaptive weighting:
\begin{equation}
    S_{\text{ranking}}
    = S_{\text{levenshtein}}
    + \lambda_{1}S_{\text{BM25}}
    + \lambda_{2}S_{\text{gnn}},
\end{equation}
where $\lambda_1$ and $\lambda_2$ balance textual and structural matching signals. For each query, we select the top-$k$ nodes with the highest $S_{\text{ranking}}$ scores as contextual evidence for answer generation.

\section{Experiments}

We conduct experiments and evaluations on different benchmark datasets to verify these research questions (RQ):
\begin{itemize}
\item  \textbf{RQ1}: How does NGM-RAG perform in different question answering tasks?
\item  \textbf{RQ2}: How do different parameter settings in NGM-RAG affect its performance?
\item  \textbf{RQ3}: How do these individual components of NGM-RAG influence its performance?
\item  \textbf{RQ4}: How does NGM-RAG balance cost and model performance?
\end{itemize}

\subsection{Experiments Setup}

\subsubsection{Datasets and Baselines}
~\newline
\noindent \textbf{Datasets.}
We evaluate NGM-RAG on three benchmarks: \textbf{HotpotQA}~\cite{yang2018hotpotqa}, \textbf{MultiHop-RAG}~\cite{tang2024multihop}, and the \textbf{UltraDomain benchmark}~\cite{qian2024memorag}. HotpotQA and MultiHop-RAG are used for multi-hop question answering, while UltraDomain is used to evaluate long-context summarization.

For UltraDomain, we follow the question generation protocol of GraphRAG~\cite{edge2024local}. For each of two domains, an LLM simulates five RAG users, each assigned five tasks. For each user-task pair, the LLM generates five questions based on textual summaries of the corpus, resulting in 125 questions per domain.

\noindent \textbf{Baselines.}
We compare NGM-RAG with representative RAG and GRAG baselines, including \textbf{NaiveRAG}~\cite{gao2023retrieval}, \textbf{GraphRAG}~\cite{edge2024local}, \textbf{LightRAG}~\cite{guo2024lightrag}, \textbf{PathRAG}~\cite{chen2025pathrag}, and \textbf{MiniRAG}~\cite{fan2025minirag}.

\subsubsection{Implementation Details}
~\newline
\noindent \textbf{Question Selection.}
We sample 500 questions from HotpotQA and 500 from MultiHop-RAG for multi-hop QA evaluation. For HotpotQA, we use the Distractor Setting, where most documents are irrelevant, requiring models to identify supporting evidence from noisy contexts.

\noindent \textbf{Evaluation Metrics.}
For HotpotQA and MultiHop-RAG, we report Exact Match (EM) and F1. EM measures whether the prediction exactly matches the ground-truth answer, while F1 evaluates word-level overlap between the prediction and reference.

For long-context summarization, we follow the pairwise LLM-based evaluation protocol of GraphRAG~\cite{edge2024local}. Given a question, target metric, and two candidate answers, the evaluator selects the better answer for each dimension and then determines the overall winner.

\noindent \textbf{Backbone LLMs.}
We use four types of backbone models for answer generation: GPT-4o-mini, Llama-3.1-8B, o1-mini, and DeepSeek-R1, including both the 8B model and the full-size API version. This selection covers both conversational and reasoning-oriented models across different scales.

\begin{table*}[h]
\centering
\resizebox{\textwidth}{!}{
\begin{tabular}{lcccccccccccccc}
\toprule
\multirow{2}{*}{\textbf{HotpotQA}} & 
\multicolumn{2}{p{6em}}{\centering\textbf{NaiveRAG}} & 
\multicolumn{2}{p{6em}}{\centering\textbf{LightRAG}} & 
\multicolumn{2}{p{6em}}{\centering\textbf{GraphRAG}} & 
\multicolumn{2}{p{6em}}{\centering\textbf{PathRAG}} & 
\multicolumn{2}{p{6em}}{\centering\textbf{MiniRAG}} &
\multicolumn{2}{p{6em}}{\centering\textbf{NGM (L)}} & 
\multicolumn{2}{p{6em}}{\centering\textbf{NGM (G)}} \\ 
\cmidrule(lr){2-3} \cmidrule(lr){4-5} \cmidrule(lr){6-7} \cmidrule(lr){8-9} \cmidrule(lr){10-11} \cmidrule(lr){12-13} \cmidrule(lr){14-15}
& \textbf{EM} & \textbf{F1} & \textbf{EM} & \textbf{F1} & \textbf{EM} & \textbf{F1} & \textbf{EM} & \textbf{F1} & \textbf{EM} & \textbf{F1} & \textbf{EM} & \textbf{F1} & \textbf{EM} & \textbf{F1} \\ 
\midrule
Llama3.1:8B       & 0.186 & 0.286 & 0.278 & 0.378 & 0.122 & 0.206 & 0.210 & 0.294 & 0.010 & 0.159 & \textbf{0.326} & \textbf{0.441} & 0.298 & 0.431  \\
GPT-4o-mini       & 0.272 & 0.376 & 0.358 & 0.505 & 0.304 & 0.403 & 0.392 & 0.516 & 0.394 & 0.507 & 0.424 & 0.564 & \textbf{0.440} & \textbf{0.572}  \\
Deepseek-R1:8B    & 0.186 & 0.283 & 0.278 & 0.378 & 0.114 & 0.195 & 0.162 & 0.242 & 0.146 & 0.189 & 0.278 & \textbf{0.414} & \textbf{0.280} & 0.411  \\
Deepseek-R1(API)  & 0.368 & 0.489 & 0.336 & 0.470 & 0.452 & 0.582 & 0.408 & 0.537 & 0.352 & 0.476 & 0.500 & 0.649 & \textbf{0.506} & \textbf{0.654}  \\
\bottomrule
\end{tabular}
}

\vspace{1em}

\resizebox{\textwidth}{!}{
\begin{tabular}{lcccccccccccccc}
\toprule
\multirow{2}{*}{\textbf{MultiHop-RAG}} & 
\multicolumn{2}{p{6em}}{\centering\textbf{NaiveRAG}} & 
\multicolumn{2}{p{6em}}{\centering\textbf{LightRAG}} & 
\multicolumn{2}{p{6em}}{\centering\textbf{GraphRAG}} & 
\multicolumn{2}{p{6em}}{\centering\textbf{PathRAG}} & 
\multicolumn{2}{p{6em}}{\centering\textbf{MiniRAG}} &
\multicolumn{2}{p{6em}}{\centering\textbf{NGM (L)}} & 
\multicolumn{2}{p{6em}}{\centering\textbf{NGM (G)}} \\ 
\cmidrule(lr){2-3} \cmidrule(lr){4-5} \cmidrule(lr){6-7} \cmidrule(lr){8-9} \cmidrule(lr){10-11} \cmidrule(lr){12-13} \cmidrule(lr){14-15}
& \textbf{EM} & \textbf{F1} & \textbf{EM} & \textbf{F1} & \textbf{EM} & \textbf{F1} & \textbf{EM} & \textbf{F1} & \textbf{EM} & \textbf{F1} & \textbf{EM} & \textbf{F1} & \textbf{EM} & \textbf{F1} \\ 
\midrule
Llama3.1:8B       & 0.554 & \textbf{0.578} & 0.482 & 0.506 & 0.320 & 0.422 & 0.474 & 0.530 & 0.286 & 0.321 & \textbf{0.556} & 0.566 & 0.536 & 0.545  \\
GPT-4o-mini       & 0.616 & 0.630 & 0.564 & 0.574 & 0.576 & 0.585 & 0.616 & 0.629 & 0.574 & 0.586 & 0.620 & 0.628 & \textbf{0.630} & \textbf{0.637}  \\
Deepseek-R1:8B    & 0.352 & 0.391 & 0.444 & 0.462 & 0.554 & 0.565 & 0.452 & 0.485 & 0.452 & 0.462 & \textbf{0.570} & \textbf{0.590} & 0.530 & 0.543  \\
Deepseek-R1(API)  & 0.534 & 0.540 & 0.470 & 0.476 & 0.548 & 0.557 & 0.490 & 0.499 & 0.540 & 0.551 & \textbf{0.600} & \textbf{0.608} & 0.578 & 0.589  \\
\bottomrule
\end{tabular}
}
\caption{The performance of NGM-RAG and baseline methods is evaluated using EM and F1. NGM (L) and NGM (G) represent two modes of NGM-RAG using homogeneous LightGCN and heterogeneous GINE, respectively.}
\label{table:main_results}
\end{table*}

\begin{table*}[h]
\small
\centering
\vspace{-0.1in}
\resizebox{\textwidth}{!}{
\begin{tabular}{@{}ccccccccccc@{}}
\toprule
\multirow{2.5}{*}{\makecell[l]{\textbf{Metrics}}} 
& \multicolumn{4}{c}{\textbf{Agriculture}}
& \multicolumn{4}{c}{\textbf{Mix}} \\ 
\cmidrule(lr){2-5} \cmidrule(lr){6-9}
& NaiveRAG & \textbf{NGM-RAG}  & LightRAG & \textbf{NGM-RAG}& NaiveRAG & \textbf{NGM-RAG}  & LightRAG & \textbf{NGM-RAG} \\
\midrule
Comprehensiveness      &4.00\%&\textbf{96.00}\%&41.60\%&\textbf{58.40}\% &20.00\%&\textbf{80.00}\%&28.80\%&\textbf{71.20}\%               \\
Diversity              &18.40\%&\textbf{81.60}\%&\textbf{77.60}\%&32.80\% &35.20\%&\textbf{64.80}\%&49.60\%&\textbf{50.40}\%              \\
Empowerment            &7.20\%&\textbf{92.80}\%&49.60\%&\textbf{50.40}\% &23.20\%&\textbf{76.80}\%&31.20\%&\textbf{68.80}\%               \\
Overall                &4.80\%&\textbf{95.20}\%&41.60\%&\textbf{58.40}\% &22.40\%&\textbf{77.60}\%&29.60\%&\textbf{70.40}\%               \\
\bottomrule
\end{tabular}
}
\caption{Comparison of dimension win rate and total win rate between NGM-RAG and baseline methods on the Agriculture and Mix datasets.}
\label{tab:compare}
\vspace{-8pt}
\end{table*}
\noindent \textbf{LLM for Basic Operations.}
For fairness, GPT-4o-mini is used for all LLM-based auxiliary operations, including graph construction, UltraDomain task and question generation, and win-rate evaluation.

\noindent \textbf{Hyperparameter Settings.}
Unless otherwise specified, we set the top-$K$ matched nodes to 5 and use $\lambda_1=\lambda_2=1.0$ for BM25 and GNN matching weights.

\subsection{Results and Performance Analysis (RQ1)}

Table~\ref{table:main_results} compares NGM-RAG, implemented with LightGCN and GINE, against baseline methods on HotpotQA and MultiHop-RAG across multiple backbone models. The results show that NGM-RAG consistently improves multi-hop question answering performance, achieving the best average results in most settings. Its gains are more evident with stronger backbone models, while the margin over LightRAG and GraphRAG becomes smaller on weaker models. Overall, NGM-RAG delivers stable improvements across datasets and model architectures, validating the robustness of graph neural networks in multi-hop reasoning and their adaptability to various model architectures.

On the UltraDomain benchmark, we evaluate methods on the Agriculture and Mix datasets using comprehensiveness, diversity, empowerment, directness, and overall win rate, as shown in Table~\ref{tab:compare}. NGM-RAG achieves the best overall win rate on both datasets and shows strong performance across multiple dimensions. Although LightRAG obtains higher diversity on Agriculture, its keyword-based retrieval may introduce redundant information which leads additional expenses or errors. In contrast, NGM-RAG better balances richness and precision, leading to more reliable long-context question answering.

Based on the above analysis, we conclude the following:

\begin{itemize}[leftmargin=*]
    \item \textbf{NGM-RAG is a versatile method for diverse tasks}: Its graph-based architecture delivers strong performance in both multi-hop question answering and long-context pairwise comparison, demonstrating broad applicability and adaptability.
    \item \textbf{NGM-RAG significantly enhances RAG task performance with stability}: Experimental results show significant improvements across various models and settings, highlighting its technical robustness and reliability in complex retrieval-augmented generation tasks.
\end{itemize}

\subsection{Parameter Analysis (RQ2)}

In this section, we conduct a series of parameter sensitivity analyses to evaluate how different configurations affect the performance of NGM-RAG.

\subsubsection{Impact of Retrieval Depth $k$}
We study the effect of retrieval depth $k$ using Llama3.1:8B. As shown in \tableref{table:k}, EM and F1 improve as $k$ increases from 1 to 5, with F1 peaking at $k=5$ (0.441). Although $k=10$ slightly improves EM to 0.330, its lower F1 suggests that excessive retrieval may introduce redundancy or noise, weakening generation quality.

\begin{table}[t]
\centering
\renewcommand{\arraystretch}{1.15}
\begin{tabular}{ccc}
\toprule
\textbf{Top-$k$ Retrieved} & \textbf{EM} & \textbf{F1} \\
\midrule
$k = 1$  & 0.240 & 0.345 \\
$k = 3$  & 0.294 & 0.388 \\
$k = 5$  & 0.326 & \textbf{0.441} \\
$k = 10$ & \textbf{0.330} & 0.432 \\
\bottomrule
\end{tabular}
\caption{HotpotQA performance of NGM-RAG under varying $k$ values.}
\label{table:k}
\end{table}

\subsubsection{Impact of Weight}
We conduct a detailed analysis of two critical hyperparameters, $\lambda_1$ and $\lambda_2$, which govern the weights assigned to text similarity and neural graph matching, respectively. To understand their impact on model performance, we vary each parameter across $\{0, 1, 5, 10\}$.

\begin{figure}[t]
\centering
\includegraphics[width=0.48\textwidth]{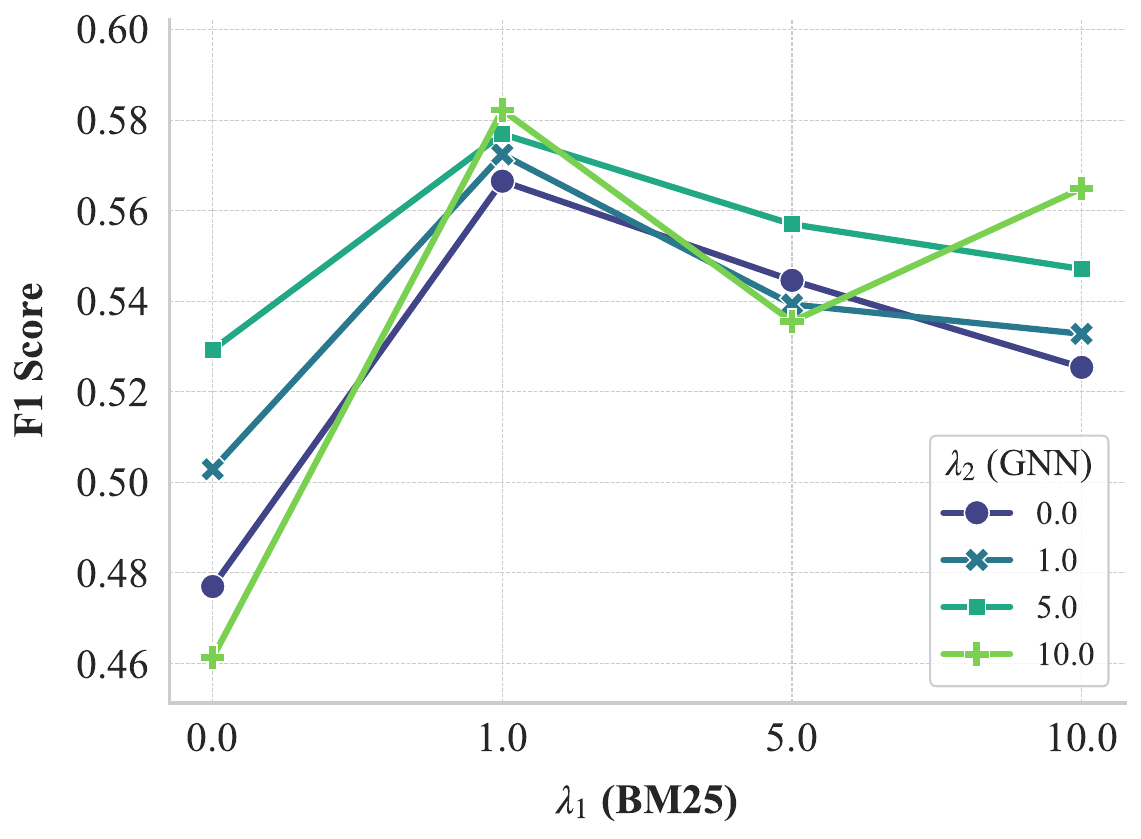}
\caption{Impact of $\lambda_1$ (BM25 weight) and $\lambda_2$ (GNN weight) on HotpotQA performance.}
\label{fig:weight}
\end{figure}

As illustrated in Figure~\ref{fig:weight}, text similarity module exhibits its optimal effectiveness at $\lambda_1 = 1$, while neural graph matching module remains relatively robust under this condition. We observe that the performance differences across nearby weight choices are generally minor, which we attribute in part to the inherent stochasticity of LLM-generated outputs rather than a consistent trend induced by the weighting scheme itself. Consequently, assigning a uniform weight ($\lambda_1 = \lambda_2 = 1$) serves as a balanced, fair, and empirically sound configuration for \ours.

\subsection{Ablation Study (RQ3)}
The matching step in our method integrates three key modules: direct matching, text similarity, and neural graph matching. To assess the individual contributions of these components, we performed ablation experiments by sequentially removing the neural graph matching and text similarity modules.

As shown in \tableref{tab:ablation}, all three modules contribute to the final performance. Ablation studies highlight the importance of all three modules for optimal performance. Direct matching ensures that nodes with identical or similar names are matched. The text similarity module, based on TF-IDF, effectively identifies relevant content by weighting term importance—crucial for accurate retrieval in multi-hop question answering. Neural graph matching captures complex semantic relationships, further enhancing matching accuracy. Together, these modules ensure robust and precise performance.

\begin{table}[t]
\centering
\renewcommand{\arraystretch}{1.15}
\resizebox{\linewidth}{!}{%
\begin{tabular}{@{}cccc@{}}
\toprule
\textbf{GNN Matching} & \textbf{Text Similarity} & \textbf{EM} & \textbf{F1} \\
\midrule
\ding{55} & \ding{55} & 0.367 & 0.477 \\
\ding{51} & \ding{55} & 0.410 & 0.529 \\
\ding{55} & \ding{51} & 0.410 & 0.545 \\
\ding{51} & \ding{51} & \textbf{0.424} & \textbf{0.564} \\
\bottomrule
\end{tabular}%
}
\caption{HotpotQA performance of NGM-RAG in ablation experiments.}
\label{tab:ablation}
\end{table}
\subsection{Cost Analysis (RQ4)}
We evaluate the efficiency of \ours~against baselines using two metrics: \textbf{Token Cost}, the number of input tokens per query, and \textbf{Inference Latency}, the end-to-end time from query input to final response.

\begin{figure}[t]
\centering
\includegraphics[width=0.48\textwidth]{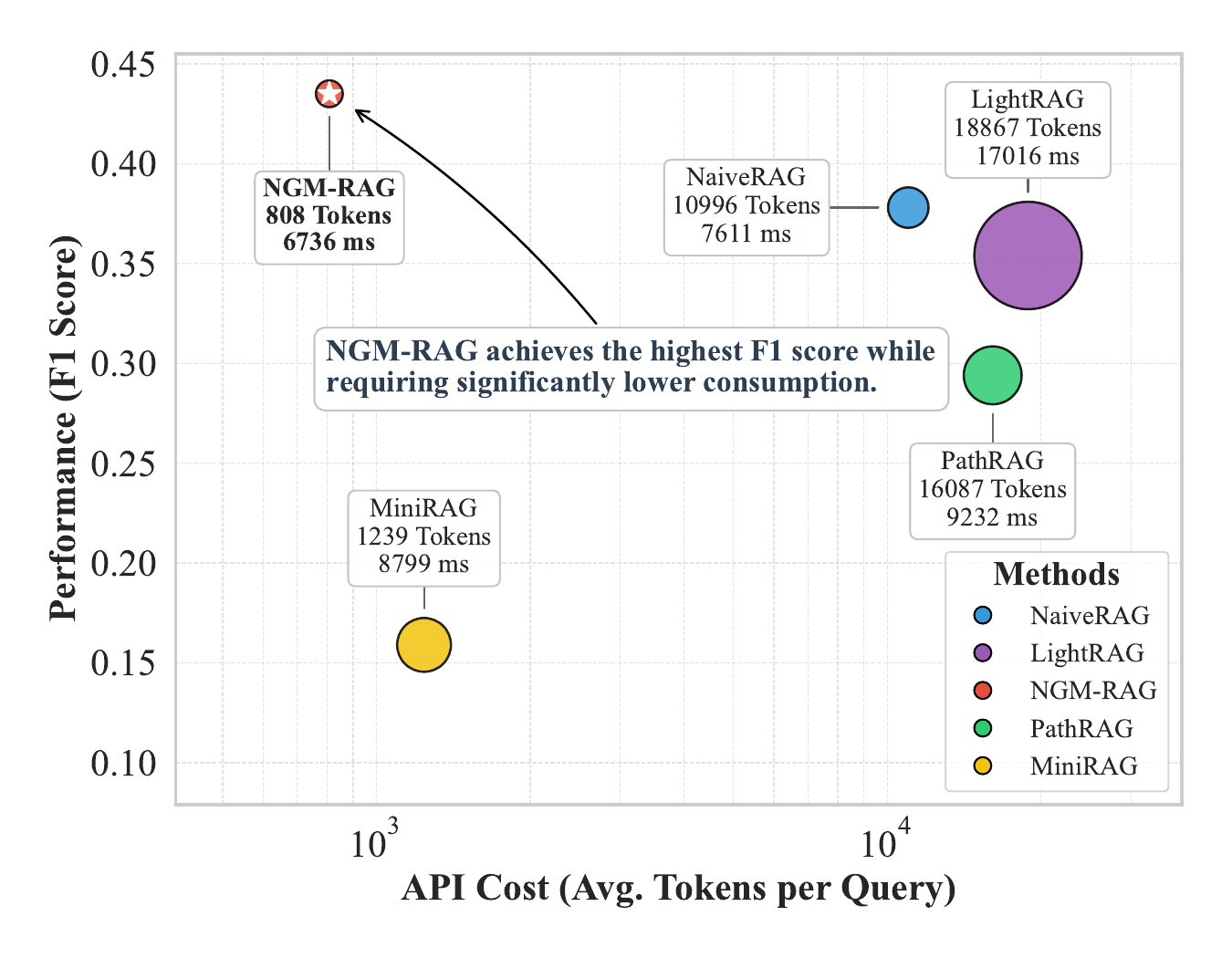}
\caption{Inference cost comparison. Bubble size indicates runtime. NGM-RAG strikes a balance between efficiency and performance.}
\label{fig:cost}
\end{figure}

As shown in \figref{fig:cost}, \ours~achieves the best balance between cost and performance. It obtains the highest F1 score while reducing the average token cost to 808.72 tokens, over 90\% lower than NaiveRAG with 10,996.07 tokens. \ours~also achieves lower latency than complex graph-based baselines, showing that neural graph matching can remove redundant context and alleviate the high-cost, high-latency bottleneck of existing GRAG methods.

\section{Conclusion}

In this paper, we present NGM-RAG, a graph matching framework for Retrieval-Augmented Generation. By integrating Levenshtein-based direct matching, BM25-based text similarity, and GNN-based neural graph matching, NGM-RAG improves retrieval precision and captures both textual and structural evidence. Experiments on multi-hop QA and long-context summarization show that NGM-RAG consistently outperforms strong RAG and GRAG baselines while reducing token cost and latency. These results demonstrate the effectiveness and efficiency of neural graph matching for complex RAG tasks.

\section*{Acknowledgements}
This work is supported by the Natural Science Foundation of China (No.62402398, No.72374173), Lab of High Confidence Embedded Software Engineering Technology and the High Performance Computing clusters at Southwest University.

\clearpage

\section*{Limitations}

While our method exhibits strong performance, it is important to acknowledge its limitations, which highlight areas for future improvement. First, the computational complexity of graph matching can significantly increase both training and inference time, particularly when applied to large-scale datasets. This results in substantial costs associated with graph construction and model training, potentially limiting its practicality in real-time or resource-constrained scenarios. Second, although NGM-RAG achieves impressive results in multi-hop reasoning and long-context summarization, its applicability to other tasks, such as open-domain question answering or dialogue systems, remains unverified. 
Further research is needed to assess its generalization capabilities across diverse domains. In addition, in pairwise comparisons of long context summaries, using a large language model for evaluation may lead to slightly unstable results. 
Finally, the current implementation relies on pre-trained language models and graph neural networks, which may inherit biases present in the training data. Addressing these limitations—through optimization of computational efficiency, exploration of broader applications, and mitigation of data biases—will be a central focus of our future work.

\section*{Potential Risks}

While our method demonstrates promising results, there are several potential risks that need to be considered. First, the reliance on pre-trained language models and graph neural networks may introduce biases present in the training data, which could lead to unfair or inaccurate outcomes in certain scenarios. Second, the computational complexity of graph matching and the integration of multiple similarity metrics may result in high resource consumption, making it less accessible for users with limited computational resources. Third, the performance of NGM-RAG is highly dependent on the quality of the input data, such as the accuracy of node names and the relevance of the retrieved documents. Noisy or incomplete data could significantly degrade the model's effectiveness. Finally, the generalization of our method to other tasks and domains remains to be thoroughly tested, and there is a risk that it may not perform as well in different contexts. Addressing these risks will be crucial for the broader adoption and reliability of NGM-RAG.

\bibliography{custom}

\clearpage
\appendix

\section{Appendix}
\subsection{Datasets}
\label{dataset}
In this section, we provide a basic introduction to the dataset used.

\textbf{HotpotQA}~\citep{yang2018hotpotqa}: A large-scale multi-hop QA dataset with 113,000 Wikipedia-based QA pairs. It requires reasoning across multiple documents and provides sentence-level supporting facts, making it ideal for evaluating explainable multi-hop reasoning.

\textbf{MultiHop-RAG}~\citep{tang2024multihop}: A benchmark with 2,556 metadata-rich queries, each requiring reasoning over 2–4 documents. It simulates real-world RAG scenarios by combining document retrieval and multi-hop reasoning.

\textbf{UltraDomain benchmark}~\citep{qian2024memorag}: A long-context QA dataset derived from 428 textbooks across 18 domains. We use the \textbf{Agriculture} and \textbf{Mix} subsets, which contain 2,017,886 tokens and 619,009 tokens respectively.  This dataset tests the robustness of the model in both technical and humanities domains through a large amount of text input.

\subsection{Baselines}
In this section, we introduce the baselines against which we compare~\ours.

\begin{itemize}
    \item \textbf{NaiveRAG}~\cite{gao2023retrieval}: Naive RAG combines LLMs with vector-based retrieval from chunked external documents. The method segments raw text into chunks and stores them in a vector database using embeddings. It retrieves top-matching text via embedding similarity.

    \item \textbf{GraphRAG}~\cite{edge2024local}: GraphRAG extracts entities and relationships from the source documents, then represents them as nodes and edges in a graph. It handles global queries across corpora by generating community summaries for related groups of entities and enhances retrieval performance.
        
    \item \textbf{LightRAG}~\cite{guo2024lightrag}: LightRAG uses a two-tier retrieval mechanism based on knowledge graphs, focuses on both low-level details and high-level concepts of queries, and supports incremental updates.

    \item \textbf{PathRAG}~\cite{chen2025pathrag}: PathRAG explicitly models relational paths within an indexing graph to jointly enhance retrieval effectiveness and generation coherence.

    \item \textbf{MiniRAG}~\cite{fan2025minirag}: MiniRAG is a lightweight RAG framework designed for environments with limited computing resources.
\end{itemize}

\label{app1}
\subsection{Prompts}
In this section, we provide a detailed overview of the prompts used throughout the various components of our pipeline. Each prompt is tailored to a specific task—ranging from knowledge graph construction to long-context answer generation and qualitative answer assessment. The prompts are designed to elicit consistent, structured, and task-relevant responses from large language models (LLMs), supporting both upstream (e.g., graph and question generation) and downstream (e.g., QA and evaluation) components in the retrieval-augmented generation (RAG) framework.
\subsubsection{Graph Construction}
\label{sec:graph-contstruction}
In this section, we describe the prompts used to extract knowledge graphs from the query and corpus for the RAG task.

Figure~\ref{fig:graph_construct} presents a prompt of entities and relations extracted from the corpus during the graph building process. The model is instructed to identify entities based on specified types, describe them, and detect meaningful relationships between them, including explanations, strength scores, and thematic keywords. The output also includes high-level content keywords and is formatted as a structured list in the specified language.

\subsubsection{Question Generation}
In this section, we describe the prompts used to generate global sensemarking questions via specific corpus.

Figure~\ref{fig:question_generate} shows a prompt designed to generate high-level sensemaking questions based on a dataset description. It instructs the model to identify five potential users, list five tasks for each user, and generate five questions per user-task pair that require a comprehensive understanding of the dataset. The prompt facilitates structured exploration of the dataset’s relevance and utility from multiple perspectives.

\subsubsection{Multihop-QA Answer Generation}
In this section, we describe the prompt used to generate concise answers for multihop question answering tasks, emphasizing direct responses without supporting context.

Figure~\ref{fig:multi_generate} presents a prompt designed for generating answers to multihop question answering (QA) tasks. The prompt instructs the model to answer queries using the provided context when relevant, or to generate answers directly when context is insufficient. It emphasizes brevity and prohibits any explanation or elaboration, requiring direct answers in either factual or yes/no form, depending on the question type. This setup ensures consistent, concise responses.

\subsubsection{Long-Context Answer Generation}
In this section, we describe the prompt used to generate comprehensive and evidence-based answers for long-context question answering, encouraging contextual summarization and the cautious integration of common knowledge.

Figure~\ref{fig:long_generate} shows a prompt designed for generating answers in long-context question answering scenarios. The prompt guides the model to summarize all relevant information from the provided context and optionally integrate common knowledge, while strictly avoiding unsupported content. This ensures that answers are both comprehensive and grounded in evidence.

\subsubsection{LLM Assessments Generation}
In this section, we describe the prompt used to judge and compare two LLM-generated answers across multiple qualitative metrics, producing structured and justified judgments to support consistent comparative analysis.

Figure~\ref{fig:judge} presents a prompt for evaluating two LLM-generated answers to the same question using four metrics: comprehensiveness, diversity, empowerment, and directness. The model is asked to select a winner for each metric with justification, then determine an overall winner. The output is structured in a standardized JSON format to support consistent and explainable comparative assessments.

\subsection{Experiment based on TF-IDF}
\label{tfidf}
In the preliminary phase of our experiments, the Text Similarity module was implemented using the classic TF-IDF algorithm. However, subsequent comparative studies demonstrated that the BM25 algorithm yielded superior performance for our specific task. Consequently, the pure TF-IDF implementation was replaced by BM25 in the final model presented in the main text. As shown in \tableref{table:tf_multihop} and \tableref{table:tf_compare}, to ensure completeness and provide a comparative baseline, the results from the initial TF-IDF experiments are included as supplementary experiments in the Appendix.

\begin{table}[h]
\centering
\resizebox{1\columnwidth}{!}{
\begin{tabular}{lcccc}
\toprule
\multirow{2}{*}{\textbf{HotpotQA}} & 
\multicolumn{2}{p{11em}}{\centering\makebox[11em]{\textbf{NGM-RAG(LightGCN)}}} & 
\multicolumn{2}{p{11em}}{\centering\makebox[11em]{\textbf{NGM-RAG(GINE)}}} \\ 
\cmidrule(lr){2-3} 
\cmidrule(lr){4-5} 
&  
\multicolumn{1}{p{5em}}{\centering\textbf{EM}} & 
\multicolumn{1}{p{5em}}{\centering\textbf{F1}} & 
\multicolumn{1}{p{5em}}{\centering\textbf{EM}} & 
\multicolumn{1}{p{5em}}{\centering\textbf{F1}} \\ 
\midrule
Llama3.1:8B       & 0.286 & 0.401 & 0.331 & 0.414  \\
GPT-4o-mini       & 0.382 & 0.526 & 0.370 & 0.503  \\
Deepseek-R1:8B    & 0.238 & 0.353 & 0.222 & 0.341  \\
Deepseek-R1:671B  & 0.440 & 0.594 & 0.430 & 0.578  \\
\bottomrule
\end{tabular}
}

\vspace{1em}

\resizebox{1\columnwidth}{!}{
\begin{tabular}{lcccc}
\toprule
\multirow{2}{*}{\textbf{MultiHop-RAG}} & 
\multicolumn{2}{p{11em}}{\centering\makebox[11em]{\textbf{NGM-RAG(LightGCN)}}} & 
\multicolumn{2}{p{11em}}{\centering\makebox[11em]{\textbf{NGM-RAG(GINE)}}} \\ 
\cmidrule(lr){2-3} 
\cmidrule(lr){4-5} 
&  
\multicolumn{1}{p{5em}}{\centering\textbf{EM}} & 
\multicolumn{1}{p{5em}}{\centering\textbf{F1}} & 
\multicolumn{1}{p{5em}}{\centering\textbf{EM}} & 
\multicolumn{1}{p{5em}}{\centering\textbf{F1}} \\ 
\midrule
Llama3.1:8B       & 0.496 & 0.522 & 0.472 & 0.490  \\
GPT-4o-mini       & 0.642 & 0.654 & 0.648 & 0.658  \\
Deepseek-R1:8B    & 0.476 & 0.499 & 0.480 & 0.498  \\
Deepseek-R1:671B  & 0.588 & 0.612 & 0.558 & 0.577  \\
\bottomrule
\end{tabular}
}
\caption{The performance of NGM-RAG variants (LightGCN and GINE) evaluated using EM and F1 on the HotpotQA and MultiHop-RAG datasets based on TF-IDF}
\label{table:tf_multihop}
\end{table}

\begin{table}[h]
\centering
\small
\resizebox{\columnwidth}{!}{
\begin{tabular}{@{}lcccc@{}}
\toprule
\multicolumn{5}{c}{\textbf{Agriculture Dataset}} \\
\midrule
\multirow{2.5}{*}{\textbf{Metrics}} & \multicolumn{2}{c}{vs NaiveRAG} & \multicolumn{2}{c}{vs LightRAG} \\
\cmidrule(lr){2-3} \cmidrule(lr){4-5}
& NaiveRAG & \textbf{NGM-RAG} & LightRAG & \textbf{NGM-RAG} \\
\midrule
Comprehensiveness & 0.00\% & \textbf{100.00}\% & 19.20\% & \textbf{80.80}\% \\
Diversity         & 8.00\% & \textbf{92.00}\%  & \textbf{67.20}\% & 32.80\% \\
Empowerment       & 0.00\% & \textbf{100.00}\% & 28.80\% & \textbf{71.20}\% \\
Overall           & 0.00\% & \textbf{100.00}\% & 23.20\% & \textbf{76.80}\% \\
\midrule
\multicolumn{5}{c}{\vspace{2pt}} \\ 
\toprule
\multicolumn{5}{c}{\textbf{Mix Dataset}} \\
\midrule
\multirow{2.5}{*}{\textbf{Metrics}} & \multicolumn{2}{c}{vs NaiveRAG} & \multicolumn{2}{c}{vs LightRAG} \\
\cmidrule(lr){2-3} \cmidrule(lr){4-5}
& NaiveRAG & \textbf{NGM-RAG} & LightRAG & \textbf{NGM-RAG} \\
\midrule
Comprehensiveness & 17.60\% & \textbf{82.40}\% & 25.60\% & \textbf{74.40}\% \\
Diversity         & 34.40\% & \textbf{65.60}\% & 48.00\% & \textbf{52.00}\% \\
Empowerment       & 19.20\% & \textbf{80.80}\% & 25.60\% & \textbf{74.40}\% \\
Overall           & 19.20\% & \textbf{80.80}\% & 25.60\% & \textbf{74.40}\% \\
\bottomrule
\end{tabular}
}
\caption{Comparison of win rates between NGM-RAG and baseline methods on the Agriculture and Mix datasets (TF-IDF based). Results are split vertically for single-column layout.}
\label{table:tf_compare}
\end{table}

\subsection{Case Study: Multihop-QA example}
This case involves a complex multi-hop question requiring identification of a country based on cues from magazine names and historical events. It evaluates each method’s ability to integrate dispersed information for accurate reasoning.

\textbf{NaiveRAG} relies on semantic vector similarity but tends to associate high-frequency terms with dominant entities like the United States, resulting in reasoning bias in long queries.

\textbf{GraphRAG} employs complex matching but may misclassify questions on smaller models, occasionally producing irrelevant response type.

\textbf{LightRAG} extracts salient keywords but lacks semantic depth, misattributing the context to Egypt due to noise sensitivity.

\textbf{NGM-RAG}, by integrating direct matching, text similarity, and neural graph matching, accurately identifies key nodes and relations (e.g., “often featured in”), enabling correct inference \textit{Israel}. Its robustness stems from a multi-module fusion architecture that balances semantic depth and structural alignment.

Overall, NGM-RAG shows clear advantages in complex multi-hop QA, with results detailed in Table~\ref{tab:case1}.

\begin{table}[h]
\centering
\begin{tabularx}{\columnwidth}{@{}l X@{}}
\toprule
\textbf{Query} & Which country, often ... to prevent major attacks? \\
\midrule
\textbf{Correct Answer} & Israel \\
\midrule
\textbf{NaiveRAG} & The United States \\
\midrule
\textbf{GraphRAG} & Yes \\
\midrule
\textbf{LightRAG} & Egypt \\
\midrule
\textbf{NGM-RAG} & Israel \\
\bottomrule
\end{tabularx}
\caption{Case Study:Example questions from the MultiHop-RAG dataset, and the answers generated by the baseline method and NGM-RAG when using Deepseek-R1 as the foundation model.}
\label{tab:case1}
\end{table}

\subsection{Case Study: Long-Context QA example}

This case focuses on a question-answering task from the Agriculture dataset, comparing the performance of LightRAG and NGM-RAG in generating summarization-based answers. The experimental results demonstrate the strengths of each method in addressing the query. LightRAG shows better performance in Diversity, providing analysis and explanations from multiple perspectives. However, NGM-RAG outperforms LightRAG in other key metrics, including Comprehensiveness, Empowerment, and Directness, resulting in a higher overall evaluation score. The detail results are shown in Table~\ref{tab:case2}.

LightRAG's strength lies in its ability to generate diverse responses by incorporating various angles of analysis. This is particularly evident in its coverage of different aspects related to the query, which enhances the richness of the generated content. However, its focus on diversity sometimes comes at the cost of depth and precision, leading to less comprehensive and direct answers. NGM-RAG, on the other hand, excels in delivering comprehensive and direct answers. By leveraging its multi-module fusion design, NGM-RAG effectively balances semantic understanding, contextual relevance, and factual accuracy. This enables it to provide answers that are not only detailed but also highly relevant and actionable, as reflected in its superior performance in Comprehensiveness, Empowerment, and Directness.

The results of this case study highlight the trade-offs between diversity and depth in summarization tasks. While LightRAG's diverse outputs are valuable for exploring multiple perspectives, NGM-RAG's ability to deliver comprehensive and direct answers makes it more suitable for applications requiring precise and actionable insights. This finding underscores the importance of selecting methods based on specific task requirements and evaluation metrics.

\begin{table*}[h]
\centering
\vspace{8pt}
\resizebox{\textwidth}{!}{
\begin{tabular}{|l|p{14cm}|}
\hline
\textbf{Query} & What role does consumer demand play in shaping beekeeping practices and sustainability efforts? (Role: Policy Maker, An official involved in creating and implementing agricultural policies.) \\ \hline
\textbf{LightRAG} & Consumer demand significantly influences beekeeping practices and sustainability efforts in several ways, reflecting both environmental considerations and market needs. 
Shaping Beekeeping Practices:
1. Honey Production Quality[...]
2. Diverse Product Offerings: [...]
Sustainability Efforts:
1. Support for Pollinator Health: [...]
2. Certification and Transparency: [...]
Conclusion:
Ultimately, the interplay between consumer demand and beekeeping practices becomes a catalyst for broader sustainability efforts. As consumers increasingly value sustainability and ethical sourcing, beekeepers are more likely to implement and expand practices that not only fulfill market demand but also enhance the health of bee ecosystems. This synergy fosters a cycle where responsible consumer choices lead to advancements in beekeeping methods, promoting both local economies and environmental sustainability. \\ \hline
\textbf{NGM-RAG} & Consumer demand significantly influences beekeeping practices and sustainability efforts by driving beekeepers to adopt methods that align with the preferences of health-conscious and environmentally aware consumers. As awareness of the declining honeybee populations and the importance of pollinators for food production rises, consumers increasingly seek products that reflect sustainable and ethical practices.
1. Healthier Products: [...]
2. Local Sourcing: [...]
3. Sustainable Practices: [...]
4. Education and Engagement: [...]
5. Market Trends: [...]
In summary, consumer demand acts as a powerful catalyst for beekeeping practices, encouraging sustainability and the adoption of methods that support environmental health, bee welfare, and the broader sustainability movement.\\ \hline
\textbf{LLM Decision} & 
(Answer1: NGM-RAG Answer2: LightRAG)

 \textbf{Comprehensiveness Winner}:Answer 1 
 
 \textbf{Explanation}: Answer 1 provides a detailed breakdown of how consumer demand influences various aspects of beekeeping practices, including health, local sourcing, sustainable practices, education, and market trends. It covers a wider range of factors and implications compared to Answer 2, which, while informative, is less detailed in its exploration of the topic. 

 \textbf{Diversity Winner}:Answer 2
 
 \textbf{Explanation}: Answer 2 presents a broader range of perspectives by discussing not only the quality of honey production but also the diversification of products and the importance of certifications. It highlights different aspects of consumer demand and its impact on beekeeping, making it more diverse in its insights compared to Answer 1.

 \textbf{Empowerment Winner}:Answer 1
 
 \textbf{Explanation}: Answer 1 empowers the reader by providing specific examples of how consumer demand shapes beekeeping practices and sustainability efforts. It emphasizes the role of education and community engagement, which can help readers understand how they can contribute to sustainable practices and make informed choices.

 \textbf{Overall Winner}:Answer 1
 
 \textbf{Explanation}: Overall, Answer 1 is the winner as it excels in comprehensiveness and empowerment, providing a detailed and clear response that directly addresses the question. While Answer 2 offers valuable diversity, it does not match the depth and clarity of Answer 1, making the latter the more effective answer.

\\ \hline
\end{tabular}
}
\caption{Case Study: Example question for Agriculture dataset, with generated answers from LightRAG and NGM-RAG, as well as evaluation by LLM}
\label{tab:case2}
\vspace{8pt}
\end{table*}

\subsection{Evaluation Metric Definitions}
\label{appendix:metrics}

To support the qualitative evaluation of long-context answers, we adopt these human-aligned metrics introduced in GraphRAG. Below are the definitions used during model assessment:

\begin{itemize}
    \item \textbf{Comprehensiveness}: Measures how thoroughly the answer addresses all aspects of the question, reflecting the level of detail and coverage provided.
    
    \item \textbf{Diversity}: Evaluates the richness and variety of perspectives or supporting points included in the answer, indicating the breadth of reasoning and insight.
    
    \item \textbf{Empowerment}: Assesses the answer's ability to enhance the reader's understanding and help them form informed judgements about the topic.
\end{itemize}

After calculating these metrics, the LLM judge will evaluate the overall performance and determine the overall winner.

\begin{figure*}[htbp]
\begin{PromptBox}{Prompt for entity and relationship extraction}
\textbf{Prompt:} \\

\textbf{Goal:} \\
Given a text document that is potentially relevant to this activity and a list of entity types, identify all entities of those types from the text and all relationships among the identified entities. Use \texttt{\{language\}} (default to English)as the output language.

\textbf{Steps:}
\begin{enumerate}
    \item Identify all entities. For each identified entity, extract the following information:
    \begin{itemize}
        \item \textbf{entity\_name}: Name of the entity, using the same language as the input text. If English, capitalize the name.
        \item \textbf{entity\_type}: One of the following types: [\texttt{\{entity\_types\}}]
        \item \textbf{entity\_description}: A comprehensive description of the entity's attributes and activities.
    \end{itemize}
    
    \item From the entities identified in Step 1, identify all pairs of (source\_entity, target\_entity) that are clearly related to each other. For each pair, extract the following:
    \begin{itemize}
        \item \textbf{source\_entity}: Name of the source entity.
        \item \textbf{target\_entity}: Name of the target entity.
        \item \textbf{relationship\_description}: Explanation of why the two entities are related.
        \item \textbf{relationship\_strength}: A numeric score indicating the strength of the relationship.
        \item \textbf{relationship\_keywords}: High-level keywords summarizing the nature of the relationship.
    \end{itemize}
    
    \item Identify high-level keywords that summarize the main concepts or themes of the entire text.
    
    \item Return all output in \texttt{\{language\}}, as a single list of entities and relationships. Use \#\# as the delimiter between records.

    \item At the end of the output, include the marker: <|COMPLETE|>
\end{enumerate}
\end{PromptBox}
\caption{The prompt for entity and relationship extraction during graph constructing.}
\label{fig:graph_construct}
\end{figure*}

\begin{figure*}[htbp]
\begin{PromptBox}{Prompt for global sensemaking question generation}
{\bf Prompt:} \\
Given the following description of a dataset:\\
\texttt{\{total\_description\}} \\
Please identify 5 potential users who would engage with this dataset. For each user, list 5 tasks they would perform with this dataset. Then, for each (user, task) combination, generate 5 questions that require a high-level understanding of the entire dataset.\\
Output the results in the following structure:
\begin{description}
    \item[User 1: [user description]]
    \begin{description}
        \item[Task 1: [task description]]
        \begin{description}  
            \item[Question 1:] 
            \item[Question 2:] 
            \item ...
            \item[Question 5:] 
        \end{description}
        \item[Task 2: [task description]]
        \item ...
        \item[Task 5: [task description]]
    \end{description}
\end{description}
\begin{description}
    \item[User 2: [user description]]
    \item ...
\end{description}
\begin{description}
    \item[User 5: [user description]]
    \item ...
\end{description}

\end{PromptBox}
\caption{The prompt for global sensemaking question generation.}
\label{fig:question_generate}
\end{figure*}

\begin{figure*}[htbp]
\begin{PromptBox}{Prompt for multihop-QA generating}
{\bf Prompt:} \\
You are a helpful assistant that uses provided context to answer queries.\\
If there is relevant information in the context provided, please follow it.\\
If there is really no relevant content in the relevant context, please generate it yourself.\\
But please don't explain the answer anyway, just answer. \\
Example 1: \\
Question: Who is the individual associated with the cryptocurrency industry facing a criminal trial on fraud and conspiracy charges, as reported by both The Verge and TechCrunch, and is accused by prosecutors of committing fraud for personal gain? \\
Answer(No explain): Sam Bankman-Fried \\
Example 2: Question: Were Scott Derrickson and Ed Wood of the same nationality? \\
Answer(No explain): Yes\\
If it is not a judgment type, just output the answer as in Example 1. \\
If your answer is a yes or no judgment, just output Yes and No like Example 2.\\
All in all, do not explain the answer and don't provide a note either, anyway! \\
Please output the answer, keep your answers as short as possible. (Even there is no context) \\
Context: \texttt{\{context\}} \\
Query: \texttt{\{query\}}\\
\end{PromptBox}
\caption{The prompt for multihop-QA generating.}
\label{fig:multi_generate}
\end{figure*}

\begin{figure*}[htbp]
\begin{PromptBox}{Prompt for long-context generating}
{\bf Prompt:} \\
You are a helpful assistant that uses provided context to answer queries. \\
Generate responses to user questions that summarize all information in the context provided and can incorporate any relevant common knowledge.\\
Do not include information without supporting evidence.\\
Context: \texttt{\{context\}} \\
Query: \texttt{\{query\}}\\
\end{PromptBox}
\caption{The prompt for long-context generating.}
\label{fig:long_generate}
\end{figure*}

\begin{figure*}[htbp]
\begin{PromptBox}{Prompt for LLM assessments generating}
{\bf Prompt:} \\
You will evaluate two answers to the same question based on these metrics: \textbf{Comprehensiveness}, \textbf{Diversity} and \textbf{Empowerment}.
\begin{itemize}
    \item \textbf{Comprehensiveness}: How much detail does the answer provide to cover all aspects and details of the question?
    \item \textbf{Diversity}: How varied and rich is the answer in providing different perspectives and insights on the question?
    \item \textbf{Empowerment}: How well does the answer help the reader understand and make informed judgments about the topic?
\end{itemize}
For each metric, choose the better answer (either Answer 1 or Answer 2) and explain why. Then, select an overall winner based on these four categories.

\vspace{1em}
\noindent\textbf{Here is the question:}\texttt{\{query\}}

\vspace{0.5em}
\noindent\textbf{Here are the two answers:}
\vspace{0.5em}
\noindent\textbf{Answer 1:}\texttt{\{answer1\}}
\vspace{0.5em}
\noindent\textbf{Answer 2:}\texttt{\{answer2\}}
\vspace{1em}

\noindent Evaluate both answers using the four metrics listed above and provide detailed explanations for each metric.
\vspace{1em}

\noindent Output your evaluation in the following JSON format:

\begin{verbatim}{
    "Comprehensiveness": {
        "Winner": "[Answer 1 or Answer 2]",
        "Explanation": "[Provide explanation here]"
    },
    ...
    "Overall Winner": {
        "Winner": "[Answer 1 or Answer 2]",
        "Explanation": "[Summarize why this answer is the overall winner
        based on the four metrics]"
    }
}
\end{verbatim}
\end{PromptBox}
\caption{The prompt for LLM assessments generating.}
\label{fig:judge}
\end{figure*}
\end{document}